\newcommand{\ivr}{r}
\newcommand{\ivp}{\phi}
\newcommand{\ivz}{z}
\newcommand{\ar}{\bar{\ivr}}
\newcommand{\ivro}{\ar}
\newcommand{\ds}{d\sigma}
\newcommand{\fx}{F}
\newcommand{\R}{H}
\newcommand{\M}{{\rm m}}
\newcommand{\T}{{\rm t}}
\newcommand{\ev}{\lambda}	
\newcommand{\energy}{{\cal F}}
\newcommand{\fant}{\phantom}
\newcommand{\Nt}{N_{\rm t}}
\newcommand{\nt}{n_{\rm t}}
\newcommand{\zt}{\zeta_{\rm t}}
\newcommand{\Nnc}{N}
\newcommand{\Nm}{N_{\rm m}}
\newcommand{\nm}{n_{\rm m}}
\newcommand{\zm}{\zeta_{\rm m}}
\documentstyle[aps,multicol]{revtex}
\input epsf
\begin{document}
\draft
\title{Multiply-connected Bose-Einstein condensed alkali gases: \\
		Current-carrying states and their decay
}
\author{
Erich J.~Mueller\rlap,\cite{REF:EJM} 
Paul M.~Goldbart\cite{REF:PMG} 
and 
Yuli Lyanda-Geller\cite{REF:YLG}}
\address{
Department of Physics, 
University of Illinois at Urbana-Champaign, 
1110 West Green Street, 
Urbana, IL 61801-3080, USA}
  \date{P-97-11-033-i; October 30, 1997}
\maketitle
\begin{abstract}
The ability to support metastable current-carrying states in 
multiply-connected settings is one of the prime signatures of 
superfluidity.  Such states are investigated theoretically for the 
case of trapped Bose condensed alkali gases, particularly 
with regard to the rate at which they decay via thermal fluctuations.
The lifetimes of metastable currents can be either longer or shorter 
than experimental time-scales.  A scheme for the experimental 
detection of metastable states is sketched.
\end{abstract}
\pacs{PACS numbers:  03.75.Fi, 05.30.Jp, 67.40.Fd}
%
\begin{multicols}{2}
\narrowtext
Multiply-connected superfluid and superconducting systems can support
states in which a persistent macroscopic particle-current flows.  
While not truly eternal, these states can have
extraordinarily long life-times, their decay requiring the occurrence of
certain relatively infrequent but nevertheless topologically accessible
(quantum or thermal) collective 
fluctuations~\cite{REF:history,REF:langer,REF:McCumber,REF:Tinkham}.
With the many considerable successes and rapid progress in the 
experimental exploration of 
Bose-Einstein condensed (BEC) alkali gas systems~\cite{REF:experiments},
it seems reasonable to anticipate that 
multiply-connected settings for BEC will soon become available,
thus allowing superfluid properties such as persistent currents
to be sought.
The purpose of the present Paper is to address, theoretically, the 
ability of BEC alkali gas systems in multiply-connected 
settings to support metastable current-carrying states, and to 
address the stability and decay of such states via thermal fluctuations.
Complementary work by Rokhsar~\cite{REF:Rokhsar} addresses 
related questions regarding the creation of these states, and 
their  stability.

We adopt a phenomenological description in which we
characterize the state
of the BEC system by a macroscopic wavefunction $\Psi({\bf r})$, in terms of which
the condensate density $n$ and current density $\bf j$ are given by
\begin{mathletters}
\begin{eqnarray}
n({\bf r})
&=&
\left\vert\Psi({\bf r})\right\vert^{2},
\\
{\bf j}({\bf r})&=&
{\hbar\over{2im}}
\left(
 \Psi^{\ast}{\bf\nabla}\Psi
-\Psi       {\bf\nabla}\Psi^{\ast}
\right).
\end{eqnarray}
\end{mathletters}%
The free energy $\energy$ of the state is given by the Gross-Pitaevski\u\i\/  form
\begin{equation}\label{EQ:energy}
\energy=\int\! d^{3}r\,
\left\{
{\hbar^{2}\over{2m}}\vert\nabla\Psi\vert^{2}
+\left(V({\bf r})-\mu\right)\vert\Psi\vert^{2}
+{g\over{2}}\vert\Psi\vert^{4}
\right\},
\end{equation} 
where $m$ is the mass of an individual atom, $V({\bf r})$ is an effective external
potential describing the magnetic and optical confinement of the atoms, 
and $g$ ($\equiv4\pi\hbar^{2}a/m$) represents the
interatomic interaction, with $a$ being an effective scattering length.  
For the sake of simplicity we neglect any possible effects due to spin.
This description is appropriate for analyzing the behavior of a BEC system
at chemical potential $\mu$ and temperature $T$~\cite{REF:GP}.
In order that the condensate be able to undergo the free-energy 
(and angular-momentum) changing fluctuation necessary for 
current-dissipation the condensate must not be isolated.  
Therefore we restrict the system to be at temperatures 
not far below the critical temperature $T_{\rm c}$, 
in which case the non-condensed atoms serve to provide an energy and 
angular momentum reservoir.

 \begin{figure}[tbh]
  \epsfxsize=3in
 \centerline{\epsfbox{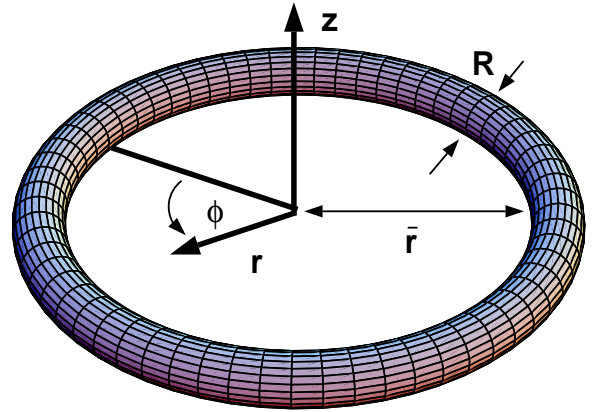}}\nobreak
 \vskip0.8cm
  \caption{
  Envisaged geometry of a trap supporting metastable current
  carrying BEC states.  The condensate healing length $\xi$
  is regarded as being small, compared with
  the circumference of the torus $L$ ($=2\pi \ar$), 
  but larger than its thickness $R$.}
  \label{FIG:geometry}
 \end{figure}
As our aim is to address multiply-connected systems, we consider trap
potentials $V({\bf r})$ that confine the gas to a cylindrically
symmetric toroidal region (Fig.~\ref{FIG:geometry}).  
Hence, $V$ depends only on 
$\ivr$ and $\ivz$, where $\{\ivr,\ivp,\ivz\}$ are the usual 
cylindrical polar coordinates.
Moreover,  we restrict our attention to
systems in which the circumference of the torus $L$ ($=2\pi \ar$) is considerably
greater than the condensate healing length $\xi$ 
[$\approx(\hbar^2/m g \bar{n})^{1/2}$, 
where $\bar{n}$ is related to the maximum particle-density], and the
thickness of the torus $R$ is comparable to or smaller than $\xi$~\cite{REF:TSG}.  
This corresponds to a regime of low condensate-density; hence, the Thomas-Fermi
approximation~\cite{REF:Baym} is not applicable.
Traps having these gross features should be achievable 
by the use of magnetic and optical forces~\cite{REF:Perhaps}. 
There are two main reasons for considering this setting:
(i)~there would be no locally-stable current-carrying states if $L$ were
comparable to or smaller than $\xi$;
(ii)~for thicker samples (i.e. $R>\xi$), the relevant 
dissipative processes by which the persistent current decays 
become
significantly more complicated (ultimately involving the nucleation of 
vortex rings)~\cite{REF:LF}.

We shall be concerned with events in which the system 
decays from some metastable 
current-carrying state $\Psi_\M$
(which is a local minimum of $\energy$) to a lower-energy 
(and typically more stable) state via a thermal fluctuation~\cite{REF:quantum}.
The current decays through a dissipative process 
during which the condensate density shrinks in magnitude over a region 
whose length is comparable to $\xi$.
Dynamically, one can envisage this process as occurring
via the passage of a vortex across the sample:  a free-energy 
barrier must be overcome for this event to occur.  The height of this
barrier $\delta F$ is given by the difference between the free energy of (the metastable
state) $\Psi_\M$ and that of the transition state $\Psi_\T$, i.e. the
lowest possible free-energy high point en-route 
through configuration space
between the initial and final metastable states. 
This thermally activated process should occur at a rate
$\omega_o e^{-\delta\energy/kT}$, 
where, as we shall discuss later, the attempt frequency $\omega_o$
does not contribute significantly to
the temperature-dependence of the rate.

In order to calculate the barrier heights,
we first identify the collection of
metastable current-carrying states $\{\Psi_\M\}$ and the relevant
states $\{\Psi_\T\}$ for transitions
between them.  Both families of states are stationary
points of $\energy$, and therefore satisfy the time-independent
Gross-Pitaevski{\u\i} equation
\begin{equation}
\frac{\delta \energy}{\delta\Psi^{\ast}}
 =-{\hbar^{2}\over{2m}}\nabla^{2}\Psi
    +\left(V({\bf r})-\mu\right)\Psi
    +g\vert\Psi\vert^{2}\Psi=0, 
\label{EQ:TIGPE}
\end{equation} 
subject to periodic boundary conditions in the  coordinate $\ivp$.

To address Eq.~(\ref{EQ:TIGPE}) we introduce a complete orthonormal
set of ``transverse'' eigenfunctions $\R_{\nu}(\ivr,\ivz)$ and the associated
energy eigenvalues $\lambda_\nu$, labeled by the
``channel'' index $\nu$, which
solve the eigenproblem
\begin{equation} 
-\frac{\hbar^2}{2 m}\left(\ivr^{-1}
\partial_\ivr
	\ivr\, \partial_\ivr
+
\partial^{2}_{\ivz}\right)\R_{\nu}
+V(\ivr,\ivz)\,\R_{\nu}=
\ev_{\nu}\R_{\nu}.
\end{equation} 
We then expand $\Psi$ in terms of these eigenfunctions:
\begin{equation} \label{EQ:expansion}
\Psi(\ivr,\ivp,\ivz)=\sum_{\nu}\fx_{\nu}(\ivp)\,\R_{\nu}(\ivr,\ivz).
\end{equation} 
By inserting this expansion into Eq.~(\ref{EQ:TIGPE})
and making use of
the orthogonality conditions for $\R_{\nu}$,
we arrive at the following
set of 
nonlinear coupled ordinary differential equations for the 
``longitudinal'' wave functions $\fx_{\nu}(\ivp)$: 
\begin{equation}\label{EQ:long}
\fx_\nu^{\prime\prime} + \alpha_\nu \fx_\nu
- \beta \sum_{\nu_1\nu_2\nu_3}\,\,\Gamma^{\nu_{\fant 0}\nu_1}_{\nu_2\nu_3}\,
\fx_{\nu_1}^\ast\,\fx_{\nu_2}^{\fant{\ast}}\,\fx_{\nu_3}^{\fant{\ast}} = C_\nu,
\end{equation}
where primes denote derivatives with respect to $\ivp$, and
the coefficients are defined by
\begin{mathletters}
\begin{eqnarray}
\alpha_\nu &\equiv& 2 m \ar^2\left(\mu-\ev_\nu\right)/\hbar^2,
\\
\beta &\equiv& 2 m \ar^2 g/\hbar^2,\\
\Gamma^{\nu_{\fant 0}\nu_1}_{\nu_2\nu_3}
&\equiv&\int\!dz\! \int\! \ivr\, d\ivr\,\, \R_{\nu}\, \R_{\nu_1}\, \R_{\nu_2}\, \R_{\nu_3}.
\end{eqnarray}
\end{mathletters}The 
terms $C_\nu$, given in footnote~\cite{REF:scatter}, are functions of the $\{\fx_{\nu}\}$ and 
are negligible when the transverse extent of the condensate $R$ is small compared
with the circumference of the torus $L$, scaling as $(R/L)^2$.

The physical condition imposed above, viz.,
that $\xi$ be much larger than the 
torus thickness $R$, 
enforces the condition
$\ev_o <\mu <\ev_{\nu\neq0}$.  
Hence, except for $\nu=0$ we have $\alpha_\nu<0$.  In practice, $\alpha_\nu$ is
expected to be quite large, scaling as $(L/R)^2$.

In the low-density limit the $\nu=0$ channel
should dominate, all other channels being occupied weakly.
We incorporate this notion by introducing a
book-keeping parameter $\Lambda$ into Eq.~(\ref{EQ:long}):
for $\nu \neq 0$ we make the replacement $\beta \to \Lambda \beta$.
It can be verified that the non-leading terms in $\Lambda$ may be neglected
for the present purposes.  (Incorporating their effects is straightforward, if
tedious.)
Similarly, one can also incorporate the effects of the $C_\nu[{\fx}]$.
Hence, we find that the 
relevant states $\{\Psi_\M\}$ and $\{\Psi_\T\}$ can adequately be described
in terms of $\fx_o$.

Within the approximation scheme just outlined, 
the uniform current-carrying states 
have the form $\Psi_\M = f_\M\, e^{iS_\M} \R_o$, with
\begin{mathletters}\label{EQ:meta}
\begin{eqnarray}
f_\M^2 &=& {\Nm}/{2\pi}
                 = (\alpha-\nm^2) /\beta\Gamma,\\
S_\M &=& \nm \ivp,
\end{eqnarray}
\end{mathletters}for 
integral $\nm$.  For the sake of brevity, 
we now write $\alpha$ and $\Gamma$ in place of $\alpha_o$ 
and $\Gamma_{oo}^{oo}$. At the stated level of approximation,
$\Nm$ is the number of condensed 
particles in the metastable state and
$\Gamma^{-1}$ is $R^2L$, i.e., the 
volume occupied by the condensate. 
By considering the second variation of $\energy$ it
can be readily shown that these states are 
local minima (and hence metastable) provided $n\zeta \leq 1$, where 
$\zm\equiv (4 \pi/\beta\Gamma \Nm)^{1/2}\approx 2 \pi \xi/ L$ 
is the dimensionless coherence length.  
(This limit on the maximum stable value of $\nm$ is 
the same as one would find 
using Landau's criterion for the critical velocity.)

The transition states $\Psi_\T=f_\T\, e^{iS_\T} \R_o$ 
are given by
\begin{mathletters}\label{EQ:soliton}
\begin{eqnarray}
f^2_\T &=&
(\Nt/2\pi)
\left(1- \Delta^2 {\rm sech}^2 (\Delta \ivp/\zt) \right),\\
f^2_\T\, \partial_\ivp S_\T &=& ({\Nt}/{2\pi})\, \nt.
\end{eqnarray}
\end{mathletters}%
Far from a region of length $\xi$, the amplitude
$f_\T$ is constant 
($f_\T^2 \sim \Nt/2\pi$) and the phase $S_\T$ winds uniformly
($S_\T \sim \nt\ivp$).
The 
coefficients in Eq.~(\ref{EQ:soliton})
appear simplest when expressed
in terms of the dimensionless coherence length
$\zt\equiv (4 \pi/\beta\Gamma \Nt)^{1/2}$: 
\begin{mathletters}
\begin{eqnarray}
\Nt/2\pi &=& ({\alpha - \nt^2})/{\beta\Gamma},\\
\nt &=& n - \pi^{-1}\cos^{-1} (\nt\,\zt),\\
\Delta^2 &=& 1-(\nt\,\zt)^2.
\end{eqnarray}
\end{mathletters}%
As $\Nm$ and $\Nt$ differ only by quantities of order
$\xi /L$, either of them may be used to characterize the number of condensed particles.
The transition states must have the property that they are saddle points of $\energy$ 
with only one direction of negative curvature.  (This unstable direction is the relevant 
reaction coordinate.)  It can readily be shown that the states
in Eq.~(\ref{EQ:soliton}) satisfy this condition as long as
$\xi>R$, or equivalently $\mu<\lambda_{\nu\neq0}$.
Thus, our approximation scheme for a BEC in a three-dimensional trap reduces the 
problem precisely to the one-dimensional problem
addressed by Little~\cite{REF:history},
Langer and  Ambegaokar~\cite{REF:langer}, and 
McCumber and Halperin~\cite{REF:McCumber}.
A useful by-product of the present approach is that it provides
a scheme for determining the intrinsic resistance of superconducting
wires clad by normal-state materials (and thus having proximity-induced
superconductivity).

Having found the relevant states, we now calculate the free-energy barrier
for dissipative fluctuations.
It can be shown that states $\Psi$ satisfying Eq.~(\ref{EQ:TIGPE})
have a free-energy
\begin{equation}
\energy = - \frac{g}{2} \sum_{\nu_0\nu_1\nu_2\nu_3}
                   \Gamma^{\nu_0\nu_1}_{\nu_2\nu_3} \int d\ivp\, 
               \fx_{\nu_0}^\ast\,\fx_{\nu_1}^\ast\,
               \fx_{\nu_2}^{\fant\ast}\,\fx_{\nu_3}^{\fant\ast}.
\end{equation}
Using this expression, along with Eq.~(\ref{EQ:meta}) and Eq.~(\ref{EQ:soliton}),
we find that 
\begin{equation}
\delta\energy = \frac{1}{2}\delta\energy_0
               \left[ \Delta \left(2+(\nt\,\zt)^2\right)
                        - 3 \nt\,\zt \cos^{-1} (\nt\,\zt) \right],
\end{equation}
where $\delta\energy_0$ is the long wavelength (i.e. $\nt\to0$) 
value of $\delta\energy$, i.e.,
\begin{equation}\label{EQ:barrier}
\delta \energy_0 =  \frac{\hbar^2}{m} 
               \left(\frac{32 \Nt^3 a}{9R^2 L^3}\right)^{1/2}.
\end{equation}

We now develop order-of-magnitude estimates for the decay rates
of metastable states via thermal fluctuations.
Let us consider $^{87}$Rb, for which the 
scattering length $a$ is 5.8nm.  
We take a harmonic trapping potential 
$V({\bf r}) = (1/2) m \omega^2 [(\ivr-\ar)^2+\ivz^2]$,
whose ground state width $\sqrt{\hbar/m\omega}$ can be
identified with the width of the condensate $R$.
To estimate $T_{\rm c}$ we consider $\Nnc$ non-interacting atoms 
in the potential $V({\bf r})$.  By virtue of the geometry (i.e. $R \ll L$) 
we can ignore the curvature of the torus, 
giving us a density of states
$\rho(E)=({4}/{3})({1}/\hbar\omega)^2({m L^2}/{2 \pi^2 \hbar^2})^{1/2} E^{3/2}$.  
Integrating this with the Bose occupation factor
reveals that $T_{\rm c}\approx 1.28 (\hbar^2/m) (\Nnc/R^4L)^{2/5}$.
For example, if we
assume that $\Nnc\approx10^6$, $\Nt\approx 2.5\times 10^4$, $R\approx1 \mu {\rm m}$,
and $L\approx100 \mu {\rm m}$ 
then we find $\delta \energy_o\,/k_{\rm B} = 3.2\mu {\rm K}$, and 
$T_{\rm c} = 0.28 \mu {\rm K}$.  The barrier height is sensitive to changes in $\Nt$, 
and can therefore be manipulated by heating or cooling the sample.

The Arrhenius formula for the decay rate in terms of the barrier
height is $\Gamma \approx \omega_o e^{-\delta \energy/kT}$.
The attempt frequency $\omega_o$ can be estimated by
using the value of the microscopic relaxation time $\tau$,
together with the assumption that
each coherence volume in the sample fluctuates 
independently~\cite{REF:history}. 
A realistic estimate for $\tau$ is the classical collision time for a dilute gas 
[i.e. $\tau^{-1} \sim \sigma n v \sim a^2 (N/V) (k_{\rm B} T/m)^{1/2} \sim 5\times 10^4$ Hz], giving 
lifetimes for the metastable states on the order of seconds.
Even beyond the limits of validity of our calculation, one expects 
$\delta\energy$ to be a monotonically increasing function of the
density.  Hence, the barriers can be extremely large at low temperatures,
allowing a continuous tuning of the metastable state lifetime 
from microseconds to times longer than the lifetime of the condensate.

We now discuss two of the issues necessary for the experimental
testing of the predictions presented in this Paper.  
We have been considering the decay of metastable states, but have not yet
addressed the issue of how to create them.
Various approaches to creating a metastable current carrying state that
rely on the superfluid properties of the condensate
have been discussed in detail 
by Rokhsar~\cite{REF:Rokhsar}.
Another technique, which does not rely on the superfluid nature of the condensate,
takes advantage of the spatial separation of the  
condensed and non-condensed atoms.
One imagines starting the 
whole system rotating (for instance by using a rotating non-axisymmetric field)
then applying a localized perturbation (such as from a sharply focused laser)
that stops the thermal atoms but leaves the condensate rotating.

The second experimental matter to be addressed is 
the detection of
metastable current-carrying states.
Perhaps the least difficult scheme would make use of
present phonon imaging techniques~\cite{REF:sound}.
The experimental configuration could be as follows:  A pulse of laser light generates a local
 rarefaction of the condensate, which then travels as two waves, one moving 
clockwise, the other counterclockwise.
By non-destructive imaging techniques one might then observe where the two waves meet, 
which gives the velocity of the metastable supercurrent.
This is only feasible if the speed of sound $c$ is comparable to the velocity $v$
with which the condensate moves around the annulus.  Linearizing
Eq.~(\ref{EQ:long}) gives
$c=(g \Gamma \Nm/2\pi m)^{1/2} \approx 1.2 {\rm mm/s}$, which is only 30 times greater than
$v=\hbar/m\ivro \approx 46 \mu{\rm m/s}$.  

\smallskip
\noindent
{\it Acknowledgments\/}: 
We
thank Gordon Baym and Tony Leggett for 
helpful discussions.  
We gratefully acknowledge support
from the Natural Sciences and Engineering Research Council of Canada
(EJM), and from the U.S.~National Science Foundation through grants
DMR94-24511 (PMG) and DMR91-57018 (YLG).


\end{multicols}

\begin{references}
\bibitem{REF:Perhaps}
For example, one can envisage creating a multiply-connected sample 
via the modification of an existing pancake-shaped 
magnetic trap~\cite{REF:experiments}, by 
puncturing it axially with a blue-detuned (and thus repulsive) laser beam.
\bibitem{REF:quantum}
In the temperature regime considered here (i.e. $T\sim T_{\rm c}$),
quantum fluctuations contribute less significantly than thermal
fluctuations, and will therefor be disregarded.

In this Paper we need not consider decay processes associated 
with quantum  fluctuations
since in the temperature region in question ($T\sim T_c$),
thermal fluctuations dominate.
\bibitem{REF:scatter}
These terms are given by
\begin{equation}
C_\nu =
\sum_{\nu_1} \alpha_{\nu_1} \Delta_{ \nu\nu_1} \fx_{\nu_1}
+\beta\,\sum_{\nu_1\nu_2\nu_3}\, \fx_{\nu_1}^\ast\fx_{\nu_2}\fx_{\nu_3}
                        D_{\nu_{\fant 0}\nu_1}^{\nu_2\nu_3},
\end{equation}
in which the quadratic coefficients $\Delta_{\nu\nu_1}$ and the quartic coefficients
$D_{\nu_{\fant 0}\nu_1}^{\nu_2\nu_3}$ are given by
\begin{mathletters}
\begin{eqnarray}
\Delta_{\nu\nu_1} &=& \ar^{-2}\int\ds\, (\ivr^2-\ar^2) \R_{\nu}\, \R_{\nu_1},\\
D_{\nu_{\fant 0}\nu_1}^{\nu_2\nu_3} &=& \ar^{-2}\int\ds\, 
                            (\ivr^2-\ar^2) \R_{\nu}\, \R_{\nu_1}\, \R_{\nu_2}\, \R_{\nu_3}.  
\end{eqnarray}
\end{mathletters}%
In these integrals the measure
$\ds$ is $\ivr\, d\ivr\, d\ivz$, with $\ivr$ ranging from 
$0$ to $\infty$ and $\ivz$ ranging over 
the whole real line.
\end{references}
\end{document}